*Article*

# Enhancing AES Using Chaos and Logistic Map-Based Key Generation Technique for Securing IoT-Based Smart Home

Ziaur Rahman [1,*], Xun Yi [1], Mustain Billah [2], Mousumi Sumi [3] and Adnan Anwar [4]

1. School Science, RMIT University, Australia; xun.yi@rmit.edu.au
2. Department of Computer Science and Engineering, Jessore University of Science and Technology, Bangladesh; mu.billah@just.edu.bd
3. Dynamic Solution Innovators (DSI), Bangladesh; mousumi.sumi@dsinnovators.com
4. Centre for Cyber Security Research and Innovation (CSRI), Deakin University, Australia; adnan.anwar@deakin.edu.au
* Correspondence: s3677291@student.rmit.edu.au; Tel.: +61-0426-117-006

**Abstract:** The Internet of Things (IoT) has brought new ways for humans and machines to communicate with each other over the internet. Though sensor-driven devices have largely eased our everyday lives, most IoT infrastructures have been suffering from security challenges. Since the emergence of IoT, lightweight block ciphers have been a better option for intelligent and sensor-based applications. When public-key infrastructure dominates worldwide, the symmetric key encipherment such as Advanced Encryption Standard (AES) shows immense prospects to sit with the smart home IoT appliances. As investigated, chaos motivated logistic map shows enormous potential to secureIoT aligned real-time data communication. The unpredictability and randomness features of the logistic map in sync with chaos-based scheduling techniques can pave the way to build a particular dynamic key propagation technique for data confidentiality, availability and integrity. After being motivated by the security prospects of AES and chaos cryptography, the paper illustrates a key scheduling technique using a 3-dimensional S-box (substitution-box). The logistic map algorithm has been incorporated to enhance security. The proposed approach has applicability for lightweight IoT devices such as smart home appliances. The work determines how seeming chaos accelerates the desired key-initiation before message transmission. The proposed model is evaluated based on the key generation delay required for the smart-home sensor devices.

## 1. Introduction

The Internet of things (IoT) has been extensively used worldwide for several purposes. The purpose includes reducing human efforts, achieving efficiency and easily understanding customer behavior, faster decision making to boost business value, etc. However, increased security concerns, such as software flaws and hacks, may cause many customers to avoid utilizing IoT devices. As shown by Figure 1, organizations in healthcare, banking, manufacturing, logistics, retail, and other industries that have already commenced to use IoT devices face severe security attacks. Attackers such as unauthorized users may establish access to connected IoT devices to misuse the network or devices. IoT devices capturing or sending information over the public channel may expose hackers if data is not protected properly. Hackers can potentially use cloud endpoints to target servers. Businesses must pay special attention to cybersecurity because of the widespread use of IoT devices. Any flaw in the system can result in a system crash or a hacker assault, which can affect huge numbers of people. For example, traffic lights may fail resulting in traffic accidents, or robbers may disable a home security system. In particular, when devices are deployed for health or human security services safety becomes more critical.

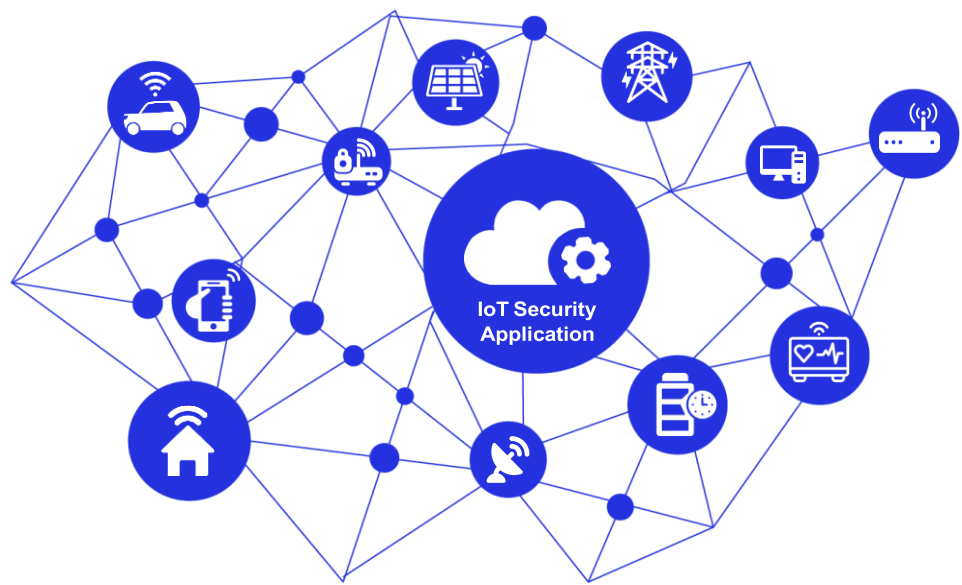

**Figure 1.** Wearables, implants, automobiles, machines, smartphones, appliances, computing systems, and other items are all capable of collecting data securely over a network. They are also able to respond to remote orders and take actions depending on collected data.

The Internet of Things (IoT) has a security concern that has successfully passed the early stages and entered the mature stage [1]. Though it was created several decades before the Advanced Encryption Standard (AES) had emerged, it has only just been available, allowing these two generations of strangers to sit together with the goal of confluencing each other [2]. Thus, thinking that both AESs are capable of securing IoT systems, particularly in terms of data integrity and secrecy, demands more justifications from trustworthy sources. On the other hand, Logistic Map (LM) with AES has evolved as a self-definable safeguard to enhance security by eliminating breaching loop-holes [3]. It has already demonstrated its capabilities in this industry by supercharging secure smart device authenticating to assure strong communication, distributed data formation, and even automated data purchase. As a result, it is conceivably expected that a developing trend of IoT utensils will be able to connect to the Internet in order to facilitate the upcoming security element of understandable and uncomplicated encryption and decryption. In the Internet of Things, a cluster head is a network device that provides reliable transmission of data by accepting data prior to actually processing and encrypting it.

Diverse complex network architecture has been evolving very rapidly. The flexibility of the network deserves special attention, as micro-shaped sensor devices have been increasingly necessary in this area of privacy and security [4]. The Advanced Encryption Standard (AES) encrypts data collected by IoT sensors, as seen in Figure 2. Although AES has demonstrated security, its key generation methodology can be breached using suitable key-breaking techniques. As a result, utilizing standard AES for crucial and real-time data protection poses a risk to data integrity. So, in our previous work [5], we combined chaos with 3DKGM [6] to overcome the drawbacks of the conventional AES. As per the evaluation conducted, the proposed technique seems to have a convincing outcome to enhance IoT security. In this paper, we have applied Chaos and Logistic Map-based Key Generation Technique for enhancing AES to secure an IoT-based smart home system.

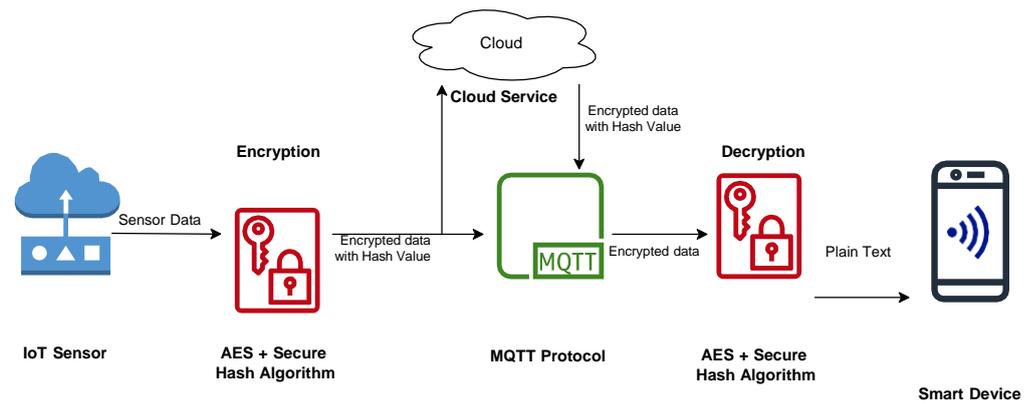

**Figure 2.** The Advanced Encryption Standard (AES) is used to safeguard data received by a smartphone from IoT sensors.

*1.1. Contributions*

The contributions claimed throughout the paper are listed below:

1. It improves the difficulty of key generation, lowering the chance of the keys being broken. The proposed key generating technique uses the 3-Dimensional Key Generation Mechanism rather than the traditional two-dimensional S-box (3DKGM). It is built on Chaos cryptography and Logistic Map to make it robust and secure.
2. The proposed method has the ability to maintain the integrity of critical IoT data and is assessed in an Iot-Based Smart-Home scenario. To adapt with the smart-home system, the sensor ought to have the requisite computing capacity.
3. The proposed method's security strength is justified by the coding-based practical and thorough evaluation when compared with similar approaches.

*1.2. Paper Organization*

The rest of the paper is organized as follows: Section 2 illustrates the background study of IoT security challenges, chaos-based *LZ*78 algorithm [7], and adaptation of AES for data integrity purposes. Related works on security vulnerabilities of IoT and chaos-based key generation are discussed in Section 3. In Section 4, proposed chaos and Logistic Map-Based Key Management Technique for IoT security is presented. Evaluation of the proposed model for ensuring security of IoT applications is discussed in Section 5. At last, the paper is concluded in Section 7.

## 2. Background

This article deals with the cybersecurity vulnerabilities of IoT applications and Chaos and Logistic Map-Based Key Generation Technique for Advanced Encryption Standards (AES) for securing IoT applications. Hence, we have divided the background discussion into three subsections:

*2.1. IoT Security Challenges*

Ensuring security in IoT environments is a challenging task due to advancement of modern technologies. Each single connection in an IoT environment introduces the possibility of security risks. A device providing more services has greater chance of being attacked. So, to protect a system, the first attempt should be to reduce the attack surface. In many cases, idle ports, which are not functioning and providing any services, are easily attacked by intruders. When devices on IoT environments interact with each others without any encryption mechanism, such plain-text-based communication has vulnerability to man-in-the-middle attacks. In that case, the intruder can access shared information. Moreover, such access to the communication link enables intruders to examine network traffic and obtain confidential and sensitive data (e.g., login credentials). The attacker can read, transmit, and modify data without the actual party's knowledge.

Even if files are encrypted, if the encryption is insufficient or incorrectly designed, there may be problems. A device, for example, might not be able to verify the legitimacy of the other party. Even though the link is encrypted, it can be intercepted by a man-in-the-middle attacker. Encryption should be used to protect sensitive data kept on the machine as well. A typical vulnerability is the lack of encryption when maintaining API tokens or credentials in plain text on a device. Other difficulties include the unintentional usage of cryptographic techniques or the use of poor encryption mechanisms. Consumer electronic devices typically contain sensitive information. The password for a wireless network is saved by devices that can be connected to it. Cameras can record both video and audio in the area where they are placed. It would be a major breach of privacy if attackers gain access to this information. IoT systems and associated services must process sensitive data with precision, security, and only with authorization of the end-user. This is valid for sensitive data generation as well as preparation.

There is no denying that security is crucial in IoT devices. The optimum technique to incorporate security in IoT-based systems, however, remains contested [8]. Cryptographic techniques are particularly prevalent among these countermeasures. The cryptographic algorithms are generic, hardware agnostic, and provide high-level robustness to IoT-based systems. AES (Advanced Encryption Standard) is an widely used cryptographic system that employs a symmetric cipher to obtain the maximum level of security. AES provides strong security features and is straightforward to implement (both software and hardware). Because of its efficient implementations, AES [9] has emerged as a viable contender for addressing the security challenges of IoT-based devices. So, in the next subsection, we discusses the role of AES for data integrity.

### 2.2. AES for Data Integrity

Even though it has been broken numerous times, the Advanced Encryption Standard (AES) is among the most ubiquitous and ever-reliable encryption algorithms. With the advancement of smart technology, new and outstanding modification approaches have been devised and used to safeguard AES as well as to improve correctness. Intruders and their unlawful access to information has now become a common phenomenon. Because of its unpredictability, chaos-based privacy has been a major problem in security research. The user may not have any prior knowledge of the initial situation, making it difficult to find the appropriate key. A tiny change in plain text or a key changes the entire result. For example, changing one bit on plain text or a key changes the result by roughly 50%. In comparison with the referred cryptosystems, chaos-based cryptosystems are more versatile for large-scale data, including audio and video. Many authors have attempted to include chaos into the current cryptosystems [10]. Other cryptographic approaches deal with the amount of integers [11], whereas chaos is addressed with actual numbers [3]. As a result, using a chaos-based method to key generation could cause the design to be safer.

### 2.3. Chaos-Based LZ78 Algorithm

This section shows how the LZ78 algorithm, which is based on chaos, can be used. In chaos-based cryptography, chaos is completely exploited. Using a one-dimensional logistic map, we may easily and safely convey information. The important properties of chaos are: it generates various intricate patterns and results in the mathematical model producing a large number of data. This information can be used to create secret keys. Stepping inside the cryptographic method requires the following operations, as indicated in Figure 3.

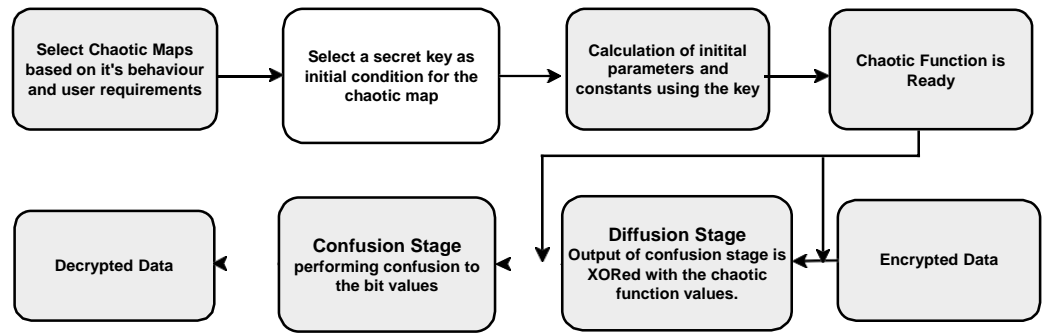

**Figure 3.** A simple Chaotic Map-based Encryption.

When we are at a period where advanced algorithms are being introduced every day, it is better to keep encryption methods simple and straightforward. The time required to encrypt and decrypt text messages should be undoubtedly the primary concern in this area of secure transmission. Long time delays during this procedure may cause the system to slow down, reducing the system's feasibility and usability. It is also worth noting that the encryption and decryption times are significantly less than those with greater probability. As a result, we would devise an algorithm that may aid in the decrease in latency for letters with a higher likelihood in a particular message. We propose that letters with multiple occurrences have the potential to generate maximum two letters in a text message. The well-known LZ78 algorithm [7] has been examined in this circumstance to reduce the time.

According to [7], LZ78 is a universally lossless compression technique. However, we use it here to shorten decryption time because, in general, the suggested approach takes longer to decrypt than it takes to encrypt. Another issue is that the decrypted output is not always correctly indexed on the decrypted transmission. As a result, we utilize LZ78 for correct indexing. It operates by calculating the likelihood of each letter in the text message. To determine the highest and lowest occurrences of a letter, a percentage is calculated.

However, factors such as IoT security challenges, use of AES for IoT data integrity, and potentiality of application of chaos-based techniques for such cases have influenced us to use chaos and Logistic Map-Based Key Generation Technique for enhancing AES to secure an IoT-based smart home.

## 3. Related Works

This article deals with the cybersecurity vulnerabilities of IoT applications and chaos and Logistic Map-Based Key Generation Technique for Advanced Encryption Standards (AES) for securing IoT applications. Hence, we have divided the review into two broad categories.

### 3.1. Related Works on Security Techniques of IoT

To safeguard the data generated by IoT devices, a variety of methods and strategies have been proposed. We covered a variety of current security algorithms and strategies for encrypting data collected from Iot devices in this section.

Thirteen (13) lightweight ciphers were analyzed in [12] in order to build a light cipher for the Internet of Things. Different factors were used to test the suggested approach. The interaction between performance characteristics and lightweight cipher component elements was investigated using association rule mining. The findings revealed some design criteria for lightweight ciphers.

Secure IoT is a lightweight encryption technique suggested by [13]. It is a symmetric key block cipher that encrypts the 64-bit block cipher using a 64-bit key. A hybrid algorithm was proposed as a solution. The algorithm has just five rounds, which uses less energy. It uses 5 keys to encrypt and decrypt because each round requires a unique key. To determine the security strength of the suggested method, many criteria were used to assess it. The technique utilized 22 bytes of memory on the ATmega 328 platform, and it took 0.188 ms to execute and 0.187 milliseconds to encrypt and decrypt.

To secure the Internet of Things, ref. [14] proposed a dynamic key approach. In this symmetric key encryption, it employed a 128-bit key that was impossible to crack by brute force. This receives data in the form of 8 bytes and outputs a fixed ciphertext of 8 bytes. The 128-bit key is divided into sixteen 8-bit subkeys. After that, shuffling operations are utilized to defend against known plaintext attacks, and finally, diffusion processes are used to create an avalanche effect.

A lightweight block cipher method named 'LiCi' was suggested in [15] that encrypts 64-bit plain text with a 128-bit key, yielding a 64-bit cipher. To encrypt the 64-bit input with a 128-bit key, LiCi consumes 1944 bytes of data and 1153 GEs (gate equivalents).The LiCi cipher uses 30 mW of power, which is less than other known approaches. It can withstand attacks from both linear and differential methods. To achieve confidentiality and data integrity, a hybrid solution [14] incorporates the benefits of both steganography and cryptography. Integrity attacks are prevented using cryptography, whereas confidentiality attacks are prevented using steganography. The model is broken into two parts. Data is transmitted between the IoT sensing element and the storage server in the first phase. The power, memory, and computing capabilities of IoT devices are all limited. Another hybrid method combines the advantages of both encryption and steganography [16]. It is quite tough for us to keep track of various mechanical lock keys in our daily lives. As a result, access control systems or electromagnetic locks have taken their place. The proposed protocol allows data to be sent between IoT devices and mobile devices.

The security of data transmission over the network was improved in [17] by combining encryption with steganography. Exchanging data such as personal data, corporate data, or state data is possible with varied levels of secrecy. As a result, the confidentiality of the data is a critical concern. To secure the data, many cryptographic techniques were developed. One of them is steganography, which is used to hide critical data. To encrypt multimedia massive data, a resource-efficient encryption approach was created in [18]. For encryption, the system does not require a second key. As a result, key distribution and updating are not required. Using the Feistel Encryption Approach, the scheme generates a key from the data. It accepts input files in the form of multi-size chunks. AES creates a ciphertext of data by encrypting it 10 times with a unique key for each round. To combine the cipher key with the cipher data, a Finely Genetic Algorithm was utilized. Decryption, on the other hand, was the inverse of the encryption strategy.

However, chaos and Logistic Map-Based Key Generation combined with AES has great potential to overcome the lacking of existing IoT security mechanisms.

*3.2. Related Works on Chaos and Logistics Map-Based Key Generation*

Several efforts were made in the area of using symmetric-key encryption algorithms to ensure safe data exchange while keeping the weight of the data low [19].

Baptista was one of the first to apply the chaotic concept [20] in the field of encryption. Other authors [21] argued that they could encrypt a message using a low-dimensional chaotic logistic map, despite the fact that it was one-dimensional. Another study advocated using a single block of short messaging text, which required fewer rounds and took longer [22].

Because of the increased sensitivity of initial condition, chaotic map is among the best data encryption algorithms [23]. Poor synchronization and severe noise problems can emerge in continuous-time chaotic nonlinear dynamics. Until now, the logistic map has been a 2-D chaotic map. However, segmented 2-D chaotic maps such as the cat map, baker map and standard map [21] have been produced in addition to the logistic map. However, the cat and baker maps have security vulnerabilities, while the standard map has not yet been extensively studied.

The LZ78 algorithm is used for lossless compression source coding. It has been used in a variety of studies. However, using it in the realm of cryptography is an uncommon occurrence. As a result, we use a logistic map in conjunction with a dynamic key generation matrix, called the 3-Dimensional Key Generation Matrix [6], to bring chaotic behavior inside. We employ it because we require more sophisticated and faster computations to secure the method and speed it up. The previous method [6] is secure with sophisticated behavior. However, encryption and decryption algorithms are concerned with both computational time as well as security. Actually, timing is a major fact to encrypt and decrypt a message. If long time latency occurs during this process, it may slow down the system. As a result, the feasibility and usability of the system is decreased. So, to keep up with modern technology and achieve both features, we need some more complicated and faster procedures.

Apart from symmetric security technique, several security approaches are found to enhance large-scale and industrial IoT security. Rahman et al. [24,25] proposed a Software Defined Netwotk (SDN)-based security technique using distributed ledger technology (DTL), which seems to be applicable for typical smart-home infrastructure. Another recent work [26,27] has been promisingly proposed: blockchain and multisignature-based certifi- cateless approaches to enhance IoT-aligned industry 4.0 security. The authors clarified the blockchain applicability for the Internet of Things (IoT) in an another paper [28]. The paper reports an obvious phenomenon: Public Key Infrastructure (PKI)-based IoT security tech- nique has several challenges that need to be addressed in an efficient and robust way. The proposed work has been motivated to bring a light-weight alternative security technique for the IoT security.

Table 1 shows the existing research works with the security features such as memory use, ability of encryption and decryption, authentication technique, data compression, algo- rithm used, efficiency, and limitations. Table 1 illustrates that the proposed 3-dimensional AES aligned with logistic map work significantly better than other state-of-the-art research work. As a result, in this research work, we have used chaos and Logistic Map-Based Key Generation Technique for enhancing AES to secure an IoT-based smart home or similar appliances with the motivation to reduce both computation and energy expenses.

**Table 1.** State-of-the-art AES improvement for IoT Security Enhancement.

| Ref. | Memory | Decrypt | Auth. | Comprs. | Alg. | Effc. | Limitations |
|---|---|---|---|---|---|---|---|
| [5,6] | √ | √ | √ | √ | 3DAES | ◯ | Time latency |
| [12,13] | √ | N | N | √ | Hybrid | ◯ | Key size |
| [14,15] | √ | √ | N | √ | LiCi | ◯ | Consuming data |
| [14,16] | √ | √ | √ | √ | Hybrid | ◯ | Complexity for IoT |
| [17,18] | √ | N | N | N | BLE | ◯ | steganography |
| [19,20] | √ | N | √ | N | AES | ◯ | Time latency |
| [21,23] | √ | √ | √ | √ | AES | ◯ | Time latency |
| [29,30] | √ | N | N | N | AES | ◯ | Not fit for IoT |
| [31,32] | √ | √ | N | N | LZ78 | ◯ | Scalability |
| [33,34] | √ | √ | √ | N | AES | ◯ | Computation high |
| [35] | √ | √ | √ | N | AES | ◯ | Computation high |

AES utilizes S-Box, which is a table-based and algorithm-based implementation. For less memory utilization, we have used the LZ78 compression method. It is also useful to speed up the communication as the size of data is decreased. For small data size, the storage of data can be performed easily. It is a necessary part of cryptography to deliver secure information and small sizes of data. So, the result of compression makes the ciphertext smaller and safer, as the ciphertext is changed after the compression occurs. Data protection with privacy is the fundamental requirement of computer security. Due to client server architecture, it is a challenging thing to give access to legal users in these complex network architectures. The efficient way of transferring data is the use of chaotic maps rather than encrypting and decrypting messages. The process of avoiding complex operations helps to create mutual authentication easily.

## 4. Proposed Chaos and Logistic Map-Based Key Management Technique for IoT Security

Because of poor power distribution, scattered nature, and lack of standardization, handling security in an IoT environment is difficult. The assumption that all cryptographic techniques are known to attackers is a risk when constructing a security system. A key is more convenient to secure against an attacker than keeping the algorithm hidden. It is also a good idea to keep the key private because it is a small bit of information. However, keeping the keys hidden is difficult due to the existence of a management system known as 'Key management'. Key management is the process of creating, modifying, altering, storing, practicing, and replacing keys. It also refers to having internal access to keys. The crucial management encompasses not only the fundamentals at the user level, but also the interactions between users. As a result, an algorithm is required to disrupt the key management procedure's internal mechanism. Instead of employing a single key, multiple keys are generated nowadays, all of which are completely reliant on one another. This could exacerbate the security breach. So, this is also a problem with today's technology. The other is a brute force attack that is limited by the length of the key. However, in this research work, we have used chaos and Logistic Map-Based Key Generation Technique for enhancing AES to secure an IoT-based smart home. So, in the following subsection, first discuss Three-Dimensional Key Generation Mechanism (3DKGM) [6]. Then, in the next subsection, we discuss the Proposed Key Generation Process.

### 4.1. Three-Dimensional Key Generation Mechanism (3DKGM)

A new key generation mechanism based on the 3DKGM matrix and S-box is developed in [1]. All of these techniques take less time for encryption and decryption than the existing AES, since they eliminate all of the time-consuming techniques. This paper, on the other hand, employs the RES approach, which is one of the most powerful algorithms ever devised. Although it guarantees less time, this article seeks to reduce the time spent in encrypting and decrypting data in comparison with other ways. Considering both security and computing speed, well-known LZ78 algorithmic properties are combined with chaos theory to achieve both qualities.

On the other hand, various logistic maps have already been developed, all of which are practical. However, in order to choose any logistic map, it must possess three characteristics: mixing capability, robust chaos, and large parameter. We use a standard logistic map after assessing all of the properties. The formula is as follows:

$$M_{p+1} = iM_p(1 - m_p) \qquad (1)$$

The range for $i$ is [0, 4], and $Mp$ is between zero and one. However, we utilize $i$ = 3.9999 in the very chaotic situation. Any encryption algorithm's head is called a key. A fundamental key generation matrix called 3DKGM (3-Dimensional Key Generation Matrix) is used in [6], which is a mixture of Latin alphabet letters, integers, and Greek values. Three keys are used here. Using 3DKGM [6], the very first key is generated.

### 4.2. Proposed Key Generation Process

Figure 4 depicts the entire key generation process. Employing the matrix inside the encryption technique is one of the most difficult challenges. At the very beginning, we declare each byte's position. After the very first key is obtained, it is necessary to note whether any byte is absent from the list. In such case, three zeros are substituted for the missing byte. From the logistic map, we now calculate the initial condition [6].

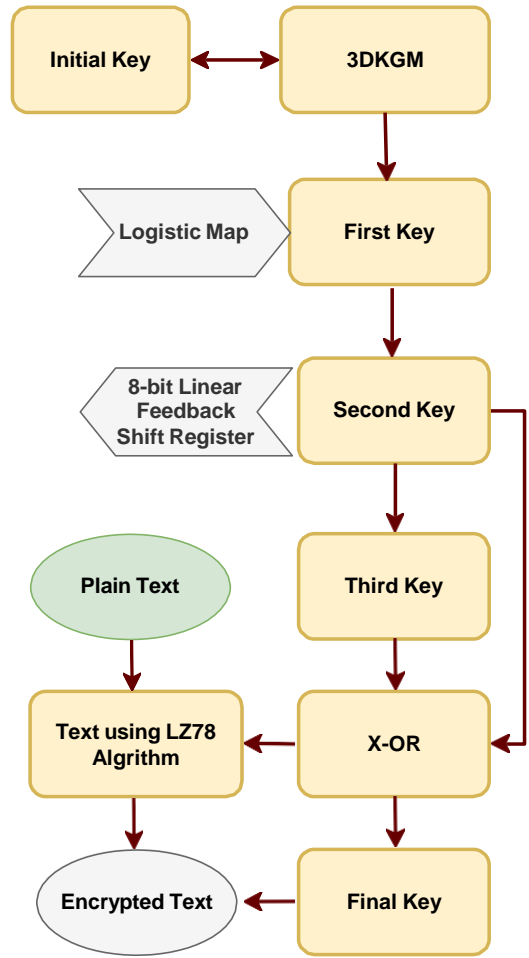

**Figure 4.** Proposed Key Generating Process.

To obtain the model parameters from the logistic map, we first select the first three blocks of the first key and transform them to binary numbers. Then, for the 2nd and 3rd keys, we execute many cycles to produce each byte. The second key and an 8-bit linear feedback shift register are required to construct the third key. Then, we perform a left shift operator on the second key and afterwards shift the bit to the right after performing an XOR (Exclusive OR) operation, resulting in the desired third key.

The very first byte of the final key is formed after the XOR function on the first, second, and third keys. Using the initial state and the following byte of the first key, several bytes are created. After continuing the cycle several times, the plaintext as well as the key are then be put through an XOR operation following the technique from [6]. As a result, it creates bytes of the key individually each time. Finally, it concatenates all of the bytes to create the final key, which is used in the encryption algorithm's further phase.

## 5. Evaluation of the Proposed Model in an IoT-Based Smart Home Environment

In this section, required computing resource and network characteristics for the implementation of this algorithm are described, and the performances of our method is evaluated and analyzed.

*5.1. Experimental Setup*

Figure 5 shows our design for an IoT-based Smart Home. The three sets of entities in our proposed smart-home architecture are: (1) consumer electronics (smoke detector, IP TV, IP camera, and smart light); (2) central node; and (3) user Interface. Wireless networking technologies are used to connect the appliances to the central controller. The smart home is controlled by the user via a user interface. IP TV, or Internet Protocol Television broadcasts via an IP network. It is the same network that is used by people to access the Internet and send emails. Because IPTV streams television over the internet, it consumes a large amount of bandwidth and data. On the other hand, an IP camera is a form of digital surveillance camera that gathers and transmits video over an IP network. Wired cameras are more reliable than Wi-Fi cameras due to various types of interference and signal degradation in wireless systems, while cameras with internal memory are more secure than cameras that record video on a cloud server. All cameras, however, can be hacked. Smart lighting is a cutting-edge method of illuminating smart homes, where a smoke detector is a device that detects smoke as a warning sign of fire. However, to implement the proposed method in the smart home, we used the Java programming language. All the results were obtained using a computer with the following specifications: Intel Core i7 CPU and 16GB RAM within the Ubuntu 18.02 operating system.

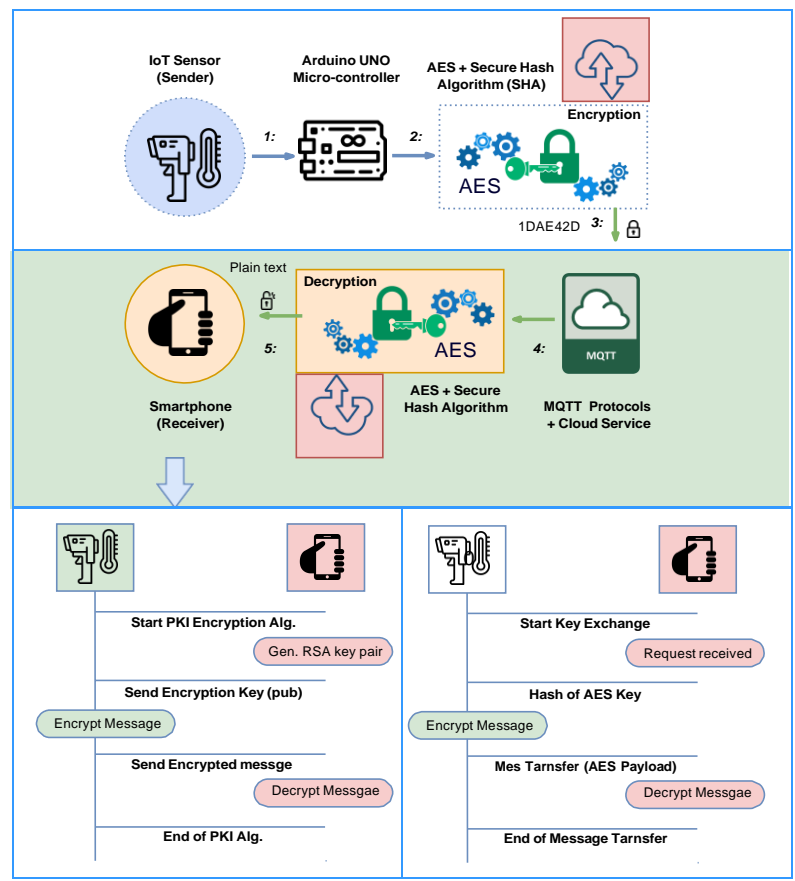

**Figure 5.** Experimental IoT Environment Setup for Evaluating the Proposed Enhanced AES Based on Chaos and LM.

*5.2. Evaluation of Required Time for Key Generation*

In this part, required time for generating keys for the proposed smart home is assessed. Our IoT-based smart home deploys four smart sensors: smoke detector, IP TV, IP camera, and smart light. However, the IP TV and IP camera consume a lot of data while transmitting signal, whereas the smoke detector and smart light transmit fewer data compared with the others. So, to evaluate the required time for the proposed key generation technique, different file sizes are used for different sensors. Required time for generating a key without and with chaos based on the existing algorithm [6] is shown in Table 2, while Table 3 illustrates required time for key generation using the proposed method. In addition, time comparison for two methods (without and with chaos) are shown in Figures 6 and 7 for smoke detector, smart light, IP TV, and IP camera, respectively.

As a result, the chaos-based approach takes less time to compute large texts than the referenced method. As a result, it provides two benefits: strength and the ability to encrypt the communication in a short amount of time.

**Table 2.** Required time for key generation using 3-dimensional key generation matrix (3DKGM) [6]. File sizes are shown in kilobit (kb) and required time is shown in millisecond (ms). For the smoke detector and smart light, the same file sizes are used, while the same file sizes are used for the IP Camera and IP TV.

| Smoke Detector | | Smart Light | | IP Camera | | IP TV | |
|---|---|---|---|---|---|---|---|
| File Size (kb) | Required Time (ms) | File Size (kb) | Required Time (ms) | File Size (kb) | Required Time (ms) | File Size (kb) | Required Time (ms) |
| 10 | 19 | 10 | 19 | 1000 | 1516 | 1000 | 1516 |
| 30 | 57 | 30 | 57 | 1500 | 1999 | 1500 | 1999 |
| 155 | 295 | 155 | 295 | 2000 | 2432 | 2000 | 2432 |
| 350 | 665 | 350 | 665 | 2500 | 2825 | 2500 | 2825 |
| 512 | 973 | 512 | 973 | 3000 | 3287 | 3000 | 3287 |

**Table 3.** Required time for key generation using the proposed method. File sizes are shown in kilobit (kb) and required time is shown in millisecond (ms). For the smoke detector and smart light, the same file sizes are used, while the same file sizes are used for the IP Camera and IP TV.

| Smoke Detector | | Smart Light | | IP Camera | | IP TV | |
|---|---|---|---|---|---|---|---|
| File Size (kb) | Required Time (ms) | File Size (kb) | Required Time (ms) | File Size (kb) | Required Time (ms) | File Size (kb) | Required Time (ms) |
| 10 | 26 | 10 | 26 | 1000 | 1489 | 1000 | 1489 |
| 30 | 67 | 30 | 67 | 1500 | 1968 | 1500 | 1968 |
| 155 | 301 | 155 | 301 | 2000 | 2356 | 2000 | 2356 |
| 350 | 671 | 350 | 671 | 2500 | 2765 | 2500 | 2765 |
| 512 | 911 | 512 | 911 | 3000 | 3200 | 3000 | 3200 |

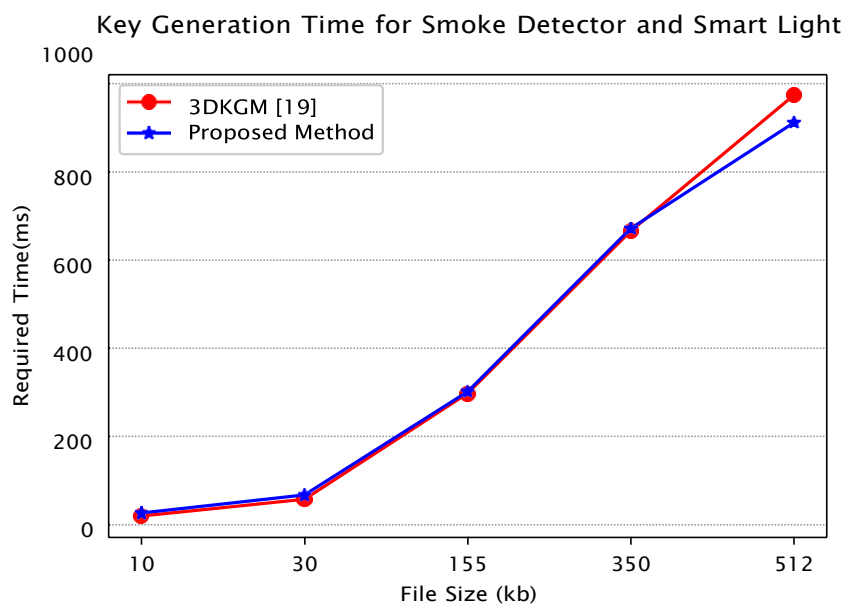

**Figure 6.** Required time for generating key using 3-dimensional key generation matrix (3DKGM)[6] and the proposed method for the case of smoke detector and smart light based on Tables 2 and 3.

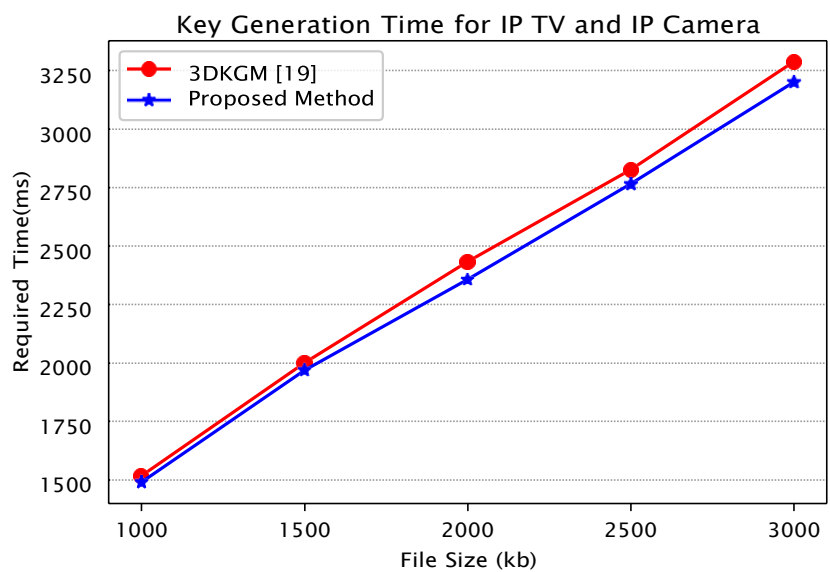

**Figure 7.** Required time for generating key using 3-dimensional key generation matrix (3DKGM) [6] and the proposed method for the case of IP TV and IP Camera based on Tables 2 and 3.

## 6. Discussion of the Effectiveness of the Proposed Model for Data Protection and Security

The proposed model has been implemented for assessing information integrity, confidentiality, and being non-repudiated in data exchange for IoT. Adding randomness and computational unpredictability makes a cryptographic solution more robust and secure. Our model has special features in encryption and decryption in terms of speed, even in building keys, and it can improve internet security.

In the proposed IoT-based smart-home scenario, each residents of the smart home has a public key, which is used to encrypt data that is sent to the smart components. Each smart component also has a private key of which the user or sender is not aware. Intruders are not be able to guess the passwords of smart appliances because of the private key. Smart components use the private key for decoding the encrypted message from the sender. The same process is followed in case of sending data from smart appliances.

However, 3DKGM [6] is dependent on the length of the initial key. Though the linear and differential attacks are theoretically impossible, this does not mean that they are immune to other types of attacks. There are various types of security attacks that require additional investigation. Today's popular saying is "be on time." So, when it comes to encrypting and decrypting a communication, timing is crucial. Because the key is a necessary and inescapable component, it is important to remember to generate it as quickly as possible. However, if it takes a short time, there is a possibility that it will be broken by intruders. To protect against all types of attack attempts, chaos seems to be the best option with the available 3DKGM algorithm for ensuring the security of a smart home.

Our experiment shows that the proposed method consumes more time with increasing file sizes, which enhances the security of the system. By increasing the number of rounds, the system becomes more secure and less prone to the attackers. With the increase in number of rounds, it requires more computational time and becomes difficult for the hacker to break the system. The generation of a key has been performed with the help of chaos and Logistic Maps. They have added more randomness and computational unpredictability. An increase in the number of rounds has brought complexity in creating keys, thus making the system complex. So, attempts to gain access to the sensor's data was totally impossible. Thus, data security, integrity, and protection of the smart-home system have been greatly improved using the proposed model.

## 7. Conclusions

When developing an IoT-based smart-home environment, it is crucial to think about security from the very beginning of the development process. However, due to the frequency of intrusions and the difficulties in reaching out to potential system vulnerabilities, guaranteeing comprehensive cybersecurity of gadgets, networks, and data in IoT contexts is difficult. It can be challenging to include comprehensive security measures in IoT applications. Aside from hardware limits, incorporating security measures may raise the pricing and development time of a system, which is not ideal for enterprises. IoT sensors require a layer of security, but they must be trusted enough to maintain data integrity. In various IoT applications, traditional AES has been shown to be vulnerable. Actually, the security of AES is dependent on the S-box and key scheduling, both of which have a substantial impact on encryption and decryption. In this paper, we devised and demonstrated an innovative key-scheduling approach that was built to encrypt massive volumes of data based on the chaotic concept linked with Logistic Map. Furthermore, we have designed an IoT-based smart-home environment to examine whether the proposed method is secure from various vulnerabilities. However, justifying the proposed scheme's continued applicability is part of the future scope. So far, the proposed technique was found to be safer for smart-home data integrity.

**Author Contributions:** Conceptualization, Z.R. and X.Y.; methodology, Z.R.; software, M.S.; validation, Z.R. and X.Y.; formal analysis, Z.R.; investigation, A.A.; resources, Z.R.; data curation, M.B.; writing—original draft preparation, Z.R.; writing—review and editing, Z.R. and M.B.; visualization, M.B.; supervision, X.Y. and A.A.; project administration, X.Y.; funding acquisition, X.Y. All authors have read and agreed to the published version of the manuscript.

**Funding:** This work was supported by the RMIT Research Stipend Scholarship (RRSS) Program. The work of Xun Yi was supported in part by the Project "Privacy-Preserving Online User Matching" under the grant ARC DP180103251. The APC was funded by the RMIT Research Stipend Scholarship (RRSS) Program.

**Institutional Review Board Statement:** Not applicable.

**Informed Consent Statement:** Informed consent was obtained from all subjects involved in the study.

**Conflicts of Interest:** The authors declare no conflict of interest.The funders had no role in the design of the study; in the collection, analyses, or interpretation of data; in the writing of the manuscript, or in the decision to publish the results.

## Abbreviations

The following abbreviations are used in this manuscript:

| | |
|---|---|
| IoT | Internet of Things |
| AES | Advanced Encryption Standard |
| LM | Logistic Map |
| 3DKGM | 3-Dimensional Key Generation Matrix |
| LZ78 | Lossless Data compression algorithms |
| GE | Gate Equivalent |
| SDN | Software Defined Network |
| DTL | Distributed Ledger Technoogy |
| PKI | Public Key Infrastructure |
| XOR | Exlusive OR |
| IP | Internet Protocol |
| TV | Televiison |
| CPU | Central Processing Unit |